\documentclass[twocolumn]{aastex62}
\usepackage{makecell}

\newcommand{\Tab}[1]{Table \ref{#1}}
\newcommand{\Fig}[1]{Figure \ref{#1}}

\received{\today}
\revised{\today}
\accepted{\today}
\submitjournal{ApJ}

\shorttitle{}
\shortauthors{Nakashima et al.}

\begin{document}

\title{X-ray Observation of a Magnetized Hot Gas Outflow in the Galactic Center Region}

\email{shinya.nakashima@riken.jp}

\author{Shinya Nakashima}
\affil{RIKEN High Energy Astrophysics Laboratory, 2-1 Hirosawa, Wako, Saitama, 351-0198, Japan}

\author{Katsuji Koyama}
\affil{Department of Physics, Graduate school of Science, Kyoto University, Kitashirakawa-Oiwake-cho, Kyoto, 606-8502, Japan}

\author{Q. Daniel Wang}
\affil{Department of Astronomy, University of Massachusetts, Amherst, MA 01003, USA}

\author{Rei Enokiya}
\affil{Department of Physics, Nagoya University, Furo-cho, Chikusa-ku, Nagoya, Aichi 464-8602, Japan}



\begin{abstract}
We report the discovery of a $1\arcdeg$ scale X-ray plume in the northern Galactic Center (GC) region observed with Suzaku.
The plume is located at ($l$, $b$) $\sim$ ($0\fdg2$, $0\fdg6$), east of the radio lobe reported by previous studies.
No significant X-ray excesses are found inside or to the west of the radio lobe.
The spectrum of the plume exhibits strong emission lines from highly ionized Mg, Si, and S that is  reproduced by a thin thermal plasma model with $kT \sim 0.7$~keV and solar metallicity. 
There is no signature of non-equilibrium ionization.
The unabsorbed surface brightness is $3\times10^{-14}$~erg~cm$^{-2}$~s$^{-1}$~arcmin$^{-2}$ in the 1.5--3.0~keV band. 
Strong interstellar absorption in the soft X-ray band indicates that the plume is not a foreground source but is at the GC distance, giving a physical size of $\sim$100~pc, a density of 0.1~cm$^{-3}$, thermal pressure of $1\times10^{-10}$~erg~cm$^{-3}$, mass of 600~$M_\sun$ and  thermal energy of $7\times10^{50}$~erg.
From the apparent association with a polarized radio emission, we propose that the X-ray plume is a magnetized hot gas outflow from the GC.
\end{abstract}




\section{Introduction}
\label{sec:int}
Outflows owing to supernovae and/or active galactic nuclei likely play an important role in the evolution of galaxies \citep[][and references therein]{2017ARA&amp;A..55...59N}.
However, the detailed physics of outflows, including the launch mechanisms, are not resolved yet.

Because of its proximity, the central hundred parsecs of the Milky Way, the Galactic Center (GC) region, is a unique laboratory for investigating the physical processes of outflows.
Indeed, the degree-scale radio lobe in the northern GC region is considered to be a relic of a past mass outflow $\sim$10~Myr ago \citep{1984Natur.310..568S,2003ApJ...582..246B,2010ApJ...708..474L}.
It might be related with the much larger scale bipolar gamma-ray bubbles found by the Fermi satellite, the so-called ``Fermi bubbles'' \citep{2010ApJ...724.1044S,2014ApJ...793...64A}.
These interpretations are supported by the discovery of past flares of the central supermassive black hole Sagittarius (Sgr) A* \citep{1996PASJ...48..249K,2009PASJ...61S.241I,2010ApJ...714..732P,2013PASJ...65...33R}. 


In the X-ray waveband, \cite{2013ApJ...773...20N} found a blob of hot gas located $\sim$$1\arcdeg$ south of the GC, whose thermal energy exceeds $10^{51}$~erg.
The hot gas is not in collisional ionization equilibrium (CIE) but in an over-ionized state on a timescale of $\sim$10$^5$~yr, suggesting that the hot gas has experienced rapid adiabatic expansion and/or photoionization.
Its origin is not clear, but it is speculated to be an outflow due to past Sgr A* flares or star-formation activities.

The above finding motivated us to survey the northern GC region.
In addition to the archival data, we made long exposure observations with the Suzaku observatory \citep{2007PASJ...59S...1M} on 2015 March.
Results from an analysis of these data are reported in this paper.

Throughout this paper, we adopt the GC distance of 8~kpc.
We used the HEASoft version 6.22 for data reduction and the Xspec version 12.9.1t for spectral analyses. 
Statistical errors are quoted at the 68\% confidence level.


\section{Observations}
\label{sec:obs}
We use the archival data of the northern GC region ($|l|<1\arcdeg$ and$-0\fdg2<b<1\fdg7$) taken during the whole life (2005--2015) of the Suzaku observatory.
We analyze the data of Suzaku/XIS \citep{2007PASJ...59S..23K}, which consisted of three front-illuminated (FI) CCDs and one back-illuminated (BI) CCD on the focal planes of the X-ray telescopes \citep{2007PASJ...59S...9S}. 
The XIS was sensitive to X-rays in the 0.3--12~keV band with the $18\arcmin \times 18\arcmin$ field-of-view. The energy resolution at 6~keV was 130--200~eV in the full width at half maximum, depending on the observation date and the CCD type \citep{2009PASJ...61S...9U}.

All the event data are reprocessed with the calibration database as of 2016 April 1st.
The cleaned events are selected using the standard screening criteria\footnote{\url{http://www.astro.isas.jaxa.jp/suzaku/process/v2changes/criteria_xis.html}}.
We then exclude additional flickering pixels reported by the instrument team\footnote{\url{http://www.astro.isas.jaxa.jp/suzaku/analysis/xis/nxb_new2/}}.
The non-X-ray background (NXB) images and spectra for each observation are estimated with the {\tt xisnxbgen} tool.

\section{Analysis and Results}
\label{sec:ana}
\subsection{X-ray Images of the Northern GC}
\begin{figure*}
\epsscale{1.2}
\plotone{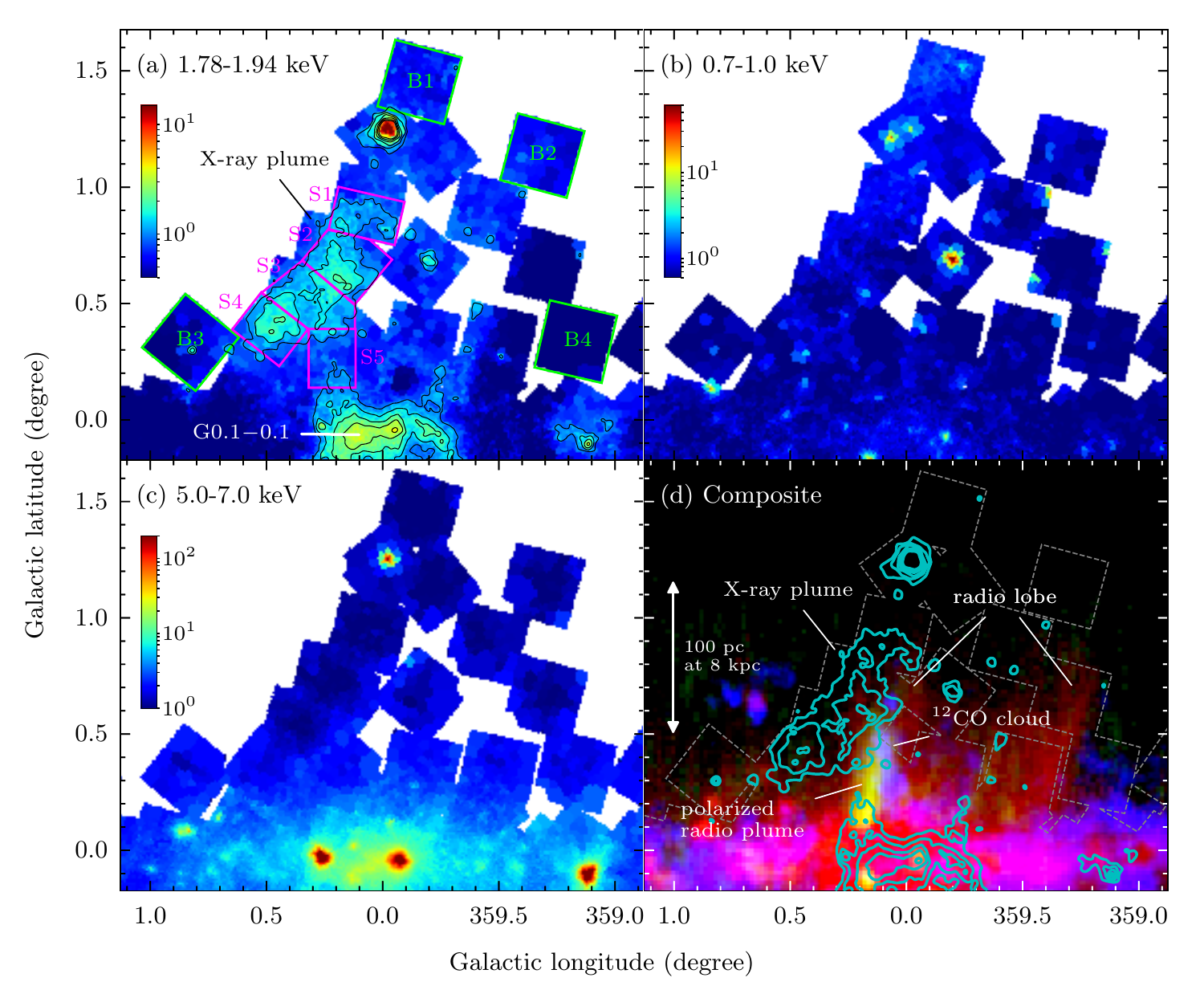}
\caption{Suzaku/XIS mosaic images of northern GC region at (a) 1.78--1.94~keV, (b) 0.7--1.0~keV, and (c) 5.0--7.0~keV.
The color bars of individual panels are in arbitrary units.
The contours in (a) are from 1.0 to 5.0 in logarithmic steps of 0.14 dex to emphasize the morphology of the X-ray plume.
The green squares (B1--B4) represent background spectral regions, whereas the magenta polygons (S1--S5) represent source spectral regions. 
(d) Composite image consisting of the Nobeyama 10 GHz continuum \citep[red;][]{1984Natur.310..568S},  Nobeyama polarized intensity at 10~GHz \cite[green;][]{1986AJ.....92..818T}, and  NANTEN $^{12}$CO ($J$ = 2$-$1) from $-3$ to 0~km~s$^{-1}$ \cite[blue;][]{2014ApJ...780...72E}. The overlaid cyan contours are the same as those in (a).  
\label{fig:image}}
\end{figure*}

\Fig{fig:image} shows  $2\arcdeg \times 2\arcdeg$ XIS mosaic images of the northern GC region in the (a) 1.78--1.94~keV, (b) 0.7--1.0~keV, and (c) 5.0--7.0~keV bands.
The first energy band includes \ion{Si}{13}~He$\alpha$, and thus, it highlights emissions from thin thermal hot gases.
The second shows foreground emissions because of strong interstellar absorption toward the GC.
The third traces the diffuse X-ray background spreading over the entire GC region \citep{2013PASJ...65...19U,2015MNRAS.453..172P}.
Differences in exposures for each observation and the vignetting effect are corrected after subtracting the NXB, and images of all the XIS sensors are co-added.
To clarify the diffuse emissions, we use an adaptive binning algorithm developed by \cite{2006MNRAS.368..497D}.

\Fig{fig:image}(a) reveals an elongated structure at ($l$, $b$) $\sim$ ($0\fdg2$, $0\fdg6$) with an angular size of $\sim$$1\arcdeg$.
This structure seems weakly connected with the intense diffuse emission
at the Sgr A region, named G0.1$-$0.1 \citep{2015MNRAS.453..172P}.
Hereafter, we refer to this structure as the ``X-ray plume''.
No corresponding structures can be seen in \Fig{fig:image}(b) or \Fig{fig:image}(c).

\subsection{Modeling of the Background Spectrum}
The diffuse X-ray background in the GC region, which is detected even at $|b| > 1\arcdeg$, is crucial to the spectral analysis of the X-ray plume.
Since the background intensity and interstellar absorption column significantly vary depending on lines of sight \citep{2013PASJ...65...19U,2016PASJ...68...59Y}, simply subtracting an off-source spectrum from the source spectrum could bias the result.
Alternatively, we construct a background spectral model using off-source regions and include it with parameter adjustments in the spectral model of the plume.

\begin{figure*}
\gridline{
	\fig{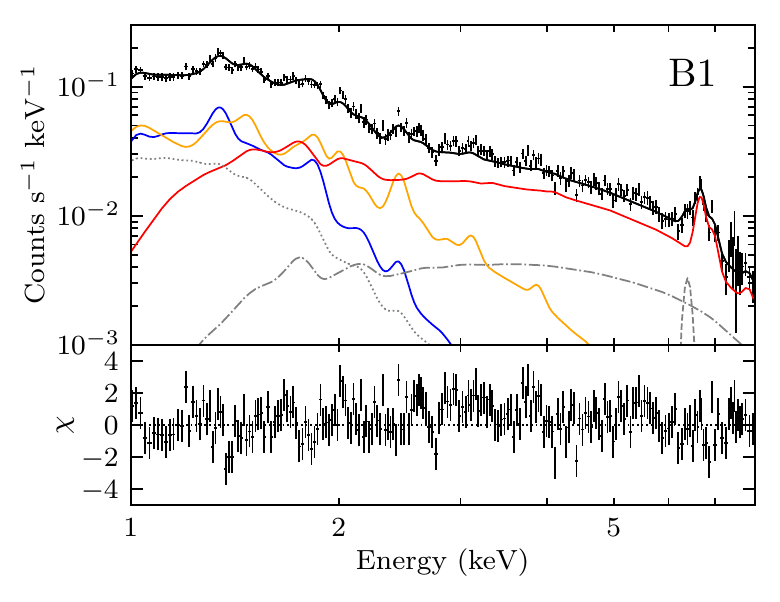}{0.45\textwidth}{}
	\fig{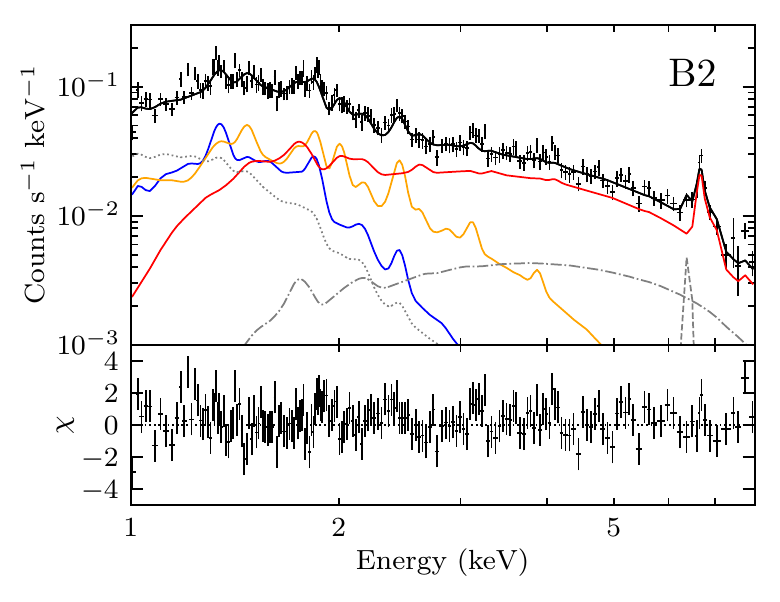}{0.45\textwidth}{}
}
\gridline{
	\fig{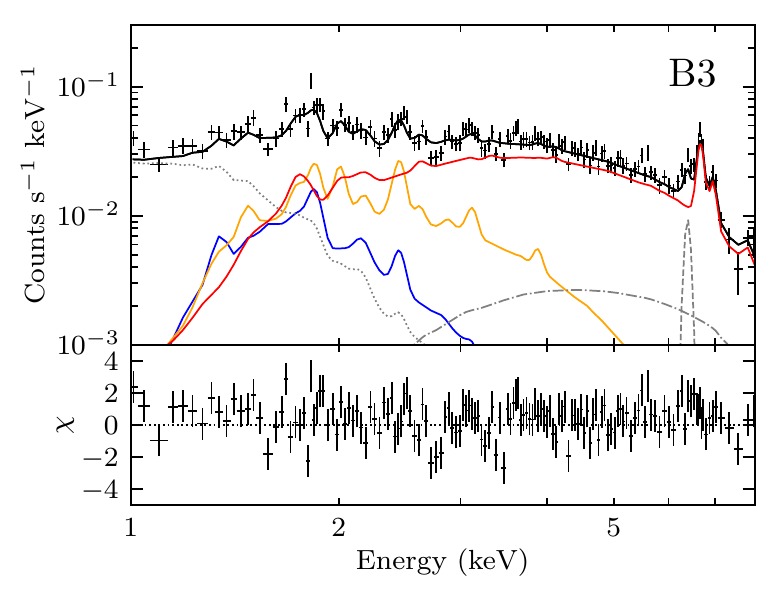}{0.45\textwidth}{}
	\fig{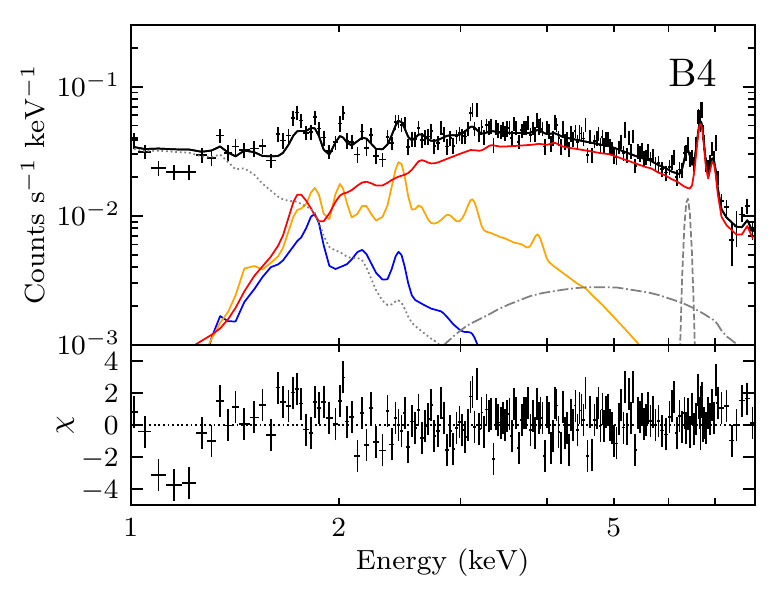}{0.45\textwidth}{}
}
\caption{XIS FI spectra of off-source regions with the model and the residuals. Spectra are re-binned for the plot. The black curves shows the total model spectra, and the other curves represent each component of the model: the low-temperature plasma (blue), middle-temperature plasma (orange), high-temperature plasma (red), \ion{Fe}{1} K$\alpha$ (dashed gray),  foreground emission (dotted gray), and CXB (dot-dashed line).\label{fig:bgdspectra}}
\end{figure*}

We use the four off-source regions, labeled B1--B4 in \Fig{fig:image}(a).
The XIS spectra of those regions, in which all the FI CCD data are co-added, are shown in \Fig{fig:bgdspectra}.
They exhibit emission lines from highly ionized Mg, Si, S, Ar, Ca, and Fe, suggesting a composite of multi-temperature hot gases.

The conventional spectral model of the GC diffuse background above $\sim$2~keV consists of two temperature ($kT\sim 1$ and 7~keV) plasma models in collisional ionization equilibrium (CIE) plus a Gaussian for \ion{Fe}{1} K$\alpha$ \citep{1996PASJ...48..249K, 2004ApJ...613..326M, 2013PASJ...65...19U,2015MNRAS.453..172P}.
However, we find that an additional low-temperature component is necessary to reproduce the $<$2~keV spectrum, especially the Mg and Si emission lines.
Therefore, we use three temperature CIE plasma models with the APEC atomic code \citep[{\tt apec} in Xspec;][]{2012ApJ...756..128F}.
The free parameters are the temperatures ($kT$), metal abundances ($Z$), and emission measures (EM).
Those for the low-, middle-, and high-temperature components are identified by subscripts L, M, and H, respectively. 
The metal abundances are reported with respect to the proto-solar composition ($Z_\sun$) of  \cite{2009LanB...4B...44L}. 
The energy centroid and width of the Gaussian are 6.4~keV and 0~eV, respectively, while its flux is allowed to vary.
The GC background emission is subject to interstellar absorption represented by the TBabs code version 2.3 \citep{2000ApJ...542..914W}, where hydrogen column density ($N_\mathrm{H, bgd}$) with the solar metal composition is a free parameter.

The spectra below $\sim$1.5~keV, especially in regions B3 and B4, are likely dominated by foreground emission because of heavy interstellar absorption toward the GC.
Following \cite{2013PASJ...65...19U}, we model the foreground emission by a single absorbed CIE plasma with an absorption column density of $5.6\times10^{21}$~cm$^{-2}$, temperature of 0.59~keV, metal abundance of 0.05~$Z_\sun$, and emission measure of $0.1$~cm$^{-6}$~pc.
Another lower temperature component ($kT = 0.09$~keV) shown in \cite{2013PASJ...65...19U} is not included in our model because it does not affect the spectrum above 1~keV.

We add the cosmic X-ray background (CXB) component, even though its contribution is minor.
The CXB spectrum is represented by a single absorbed power-law function with a photon index of 1.45 and normalization at 1~keV of $10.91$~keV~cm$^{-2}$~s$^{-1}$~sr$^{-1}$~keV$^{-1}$ \citep{2017ApJ...837...19C}. 
The chosen absorption column density is two times higher than $N_\mathrm{H, bgd}$ because the CXB comes from extragalactic sources through the Galactic disk.

The XIS spectra are binned such that each bin contains at least 20 events.
The B1--B4 spectra are simultaneously fitted with the above model to minimize the $\chi^2$ statistic.
Response matrix files and  ancillary response files are generated with the {\tt xisrmfgen} and {\tt xisximarfgen} tools and are used to convolve the models.
We assume that parameters of the plasma component and of \ion{Fe}{1} K$\alpha$ are common between regions except for the entire normalization ($f_\mathrm{bgd}$).
The normalization for B1 is fixed to 1 as a reference, while those for the other regions are allowed to vary.
The parameters of $N_\mathrm{H, bgd}$ for each region vary independently.

The best-fit parameters are tabulated in \Tab{tab:bgdfit}, and the model curves are overlaid on the data in \Fig{fig:bgdspectra}.
Our model reproduces the observed spectra with $\chi ^{2}_\nu = 1.07$.
If we add a model systematic error of 4\% to represent the instrumental calibration uncertainties, $\chi ^{2}_\nu$ decreases to 1.03 and the fit is formally acceptable with a null hypothesis probability of $>$5\%.
The temperatures and metal abundances of the middle- and high-temperature components roughly agree with those derived from \cite{2013PASJ...65...19U}.
The low-temperature component explains most of the \ion{Mg}{11}~He$\alpha$ flux and a part of the \ion{Si}{13}~He$\alpha$ flux.
$N_\mathrm{H,bgd}$ and $f_\mathrm{bgd}$ for each region increase with decreasing distance from the Galactic plane as expected.

We also fit this model to spectra of other regions in \Fig{fig:image} allowing only $N_\mathrm{H,bgd}$ and $f_\mathrm{bgd}$ to vary.
The spectra of all regions at $|b|>0\fdg3$ other than the X-ray plume region are reproduced by this background model, indicating that only the X-ray plume significantly exceeds the background emission at $|b|>0\fdg3$.
Moreover, the temperatures of the background model ($kT_\mathrm{L}$, $kT_\mathrm{M}$, and $kT_\mathrm{H}$) are different from that of the X-ray plume (see the next section). 
This suggests that the origin of the background, which is beyond the scope of this paper, is different from that of the X-ray plume.

\begin{deluxetable*}{lcccc}
\tabletypesize{\small}
\tablecaption{Background model paraemters.\label{tab:bgdfit}}
\tablehead{
\colhead{Parameters (Units)} & \colhead{B1} & \colhead{B2} & \colhead{B3} & \colhead{B4}
}
\startdata
$N_\mathrm{H,bgd}$ (10$^{22}$~cm$^{-2}$) & ${1.47}^{+0.04}_{-0.03}$ & ${2.09}^{+0.05}_{-0.04}$ & ${4.66}^{+0.08}_{-0.08}$ & ${6.48}^{+0.10}_{-0.10}$ \\
$f_\mathrm{bgd}$ & $1$ (fixed) & ${1.15}^{+0.02}_{-0.02}$ & ${2.36}^{+0.05}_{-0.05}$ & ${2.76}^{+0.05}_{-0.05}$ \\
$kT_\mathrm{L}$ (keV) & \multicolumn{4}{c}{${0.43}^{+0.01}_{-0.03}$} \\
$kT_\mathrm{M}$ (keV) & \multicolumn{4}{c}{${1.04}^{+0.04}_{-0.03}$} \\
$kT_\mathrm{H}$ (keV) & \multicolumn{4}{c}{${7.45}^{+0.24}_{-0.27}$} \\
$Z_\mathrm{L}$ ($Z_{\sun}$) & \multicolumn{4}{c}{${0.18}^{+0.04}_{-0.03}$} \\
$Z_\mathrm{M}$ ($Z_{\sun}$) & \multicolumn{4}{c}{${0.77}^{+0.22}_{-0.13}$} \\
$Z_\mathrm{H}$ ($Z_{\sun}$) & \multicolumn{4}{c}{${1.00}^{+0.06}_{-0.06}$} \\
$EM_\mathrm{L}$ (cm$^{-6}$~pc) & \multicolumn{4}{c}{${0.59}^{+0.15}_{-0.11}$} \\
$EM_\mathrm{M}$ (cm$^{-6}$~pc) & \multicolumn{4}{c}{${0.10}^{+0.02}_{-0.02}$} \\
$EM_\mathrm{H}$ (cm$^{-6}$~pc) & \multicolumn{4}{c}{${0.07}^{+0.00}_{-0.00}$} \\
\ion{Fe}{1} K$\alpha$ ($10^{-6}$ ph cm$^{-2}$ s$^{-1}$) & \multicolumn{4}{c}{${7.5}^{+0.8}_{-0.7}$} \\
\tableline
$\chi^2/\nu$ & \multicolumn{4}{c}{5092.6/4769} \\
\enddata 
\tablecomments{All the parameters but $N_\mathrm{{H,bgd}}$ and $f_\mathrm{{bgd}}$ are common to all the regions}
\end{deluxetable*}

\subsection{Spectrum of the X-ray Plume}
\begin{figure*}
\gridline{
	\fig{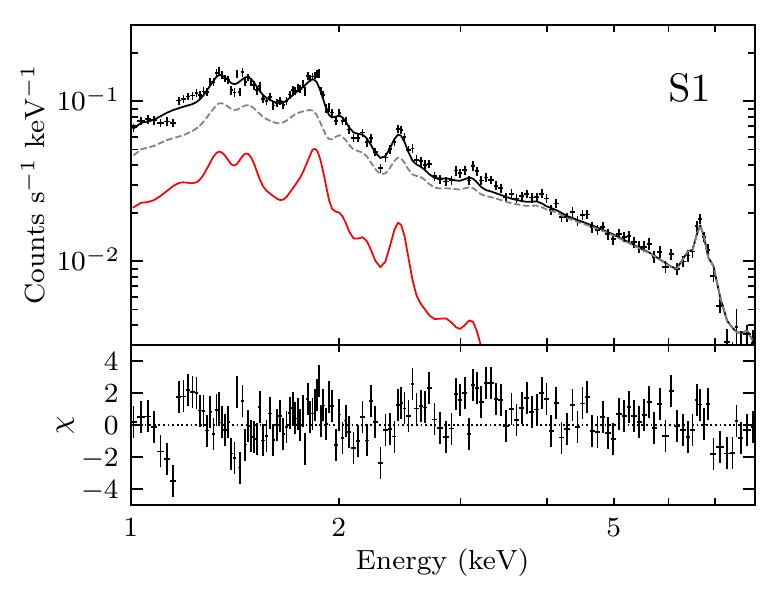}{0.45\textwidth}{}
	\fig{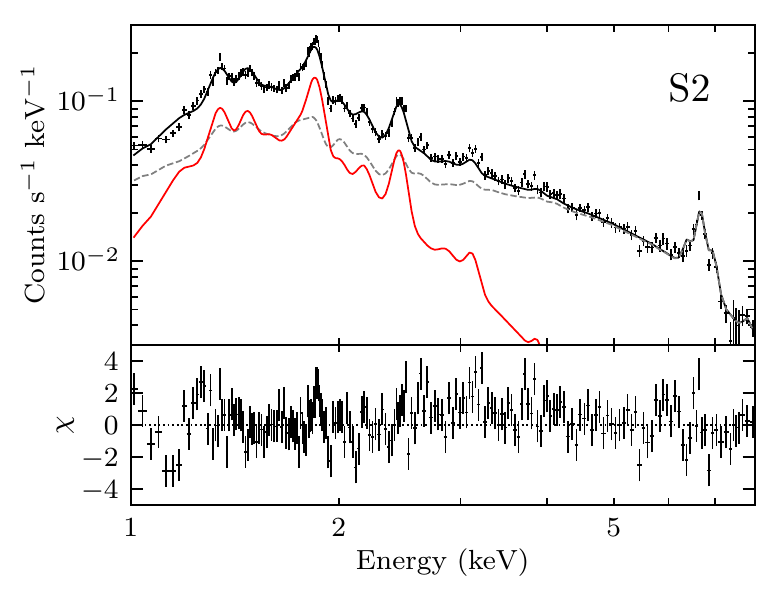}{0.45\textwidth}{}
}
\gridline{
	\fig{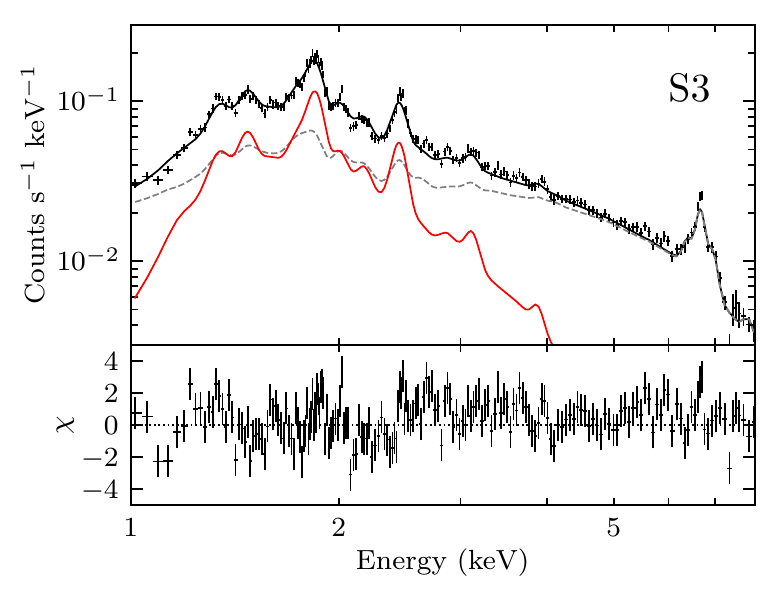}{0.45\textwidth}{}
	\fig{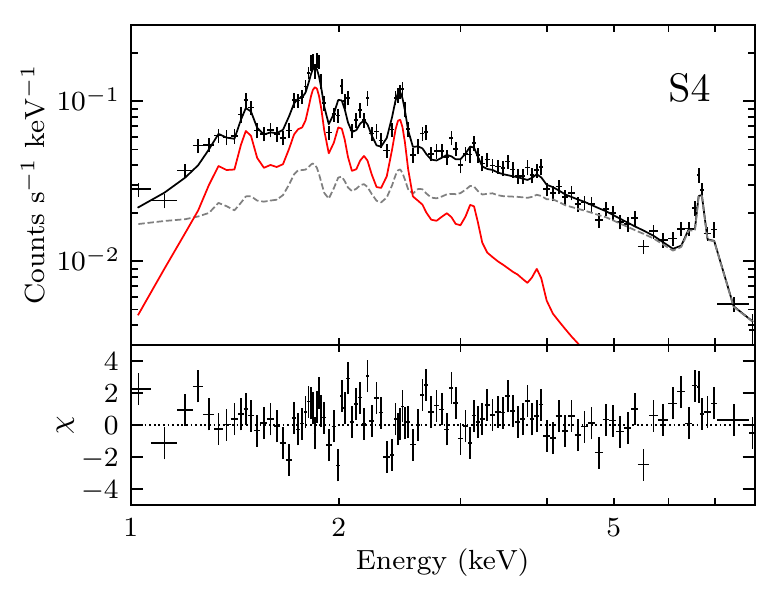}{0.45\textwidth}{}
}
\gridline{
	\fig{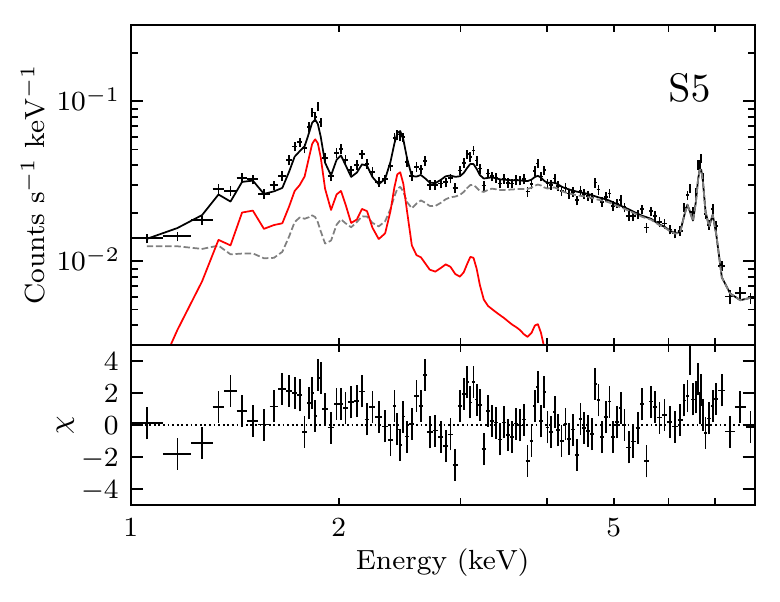}{0.45\textwidth}{}
}
\caption{XIS FI spectra of the X-ray plume regions with the model and the residuals. Spectra are re-binned for the plot. The black curves are the total model spectra while the red and dashed-gray curves represents source and background components, respectively.
\label{fig:src_spectra}}
\end{figure*}

Since the absorption column density strongly depends on the Galactic latitude as shown in the background modeling (\Tab{tab:bgdfit}), we divide the X-ray plume into five sub-regions labeled S1--S5 in \Fig{fig:image}(a). 
The derived spectra are shown in \Fig{fig:src_spectra}.
The background spectral model from the previous section does not reproduce the spectra leaving large residuals especially at Si and S emission lines.

We represent the X-ray plume spectra by adding another {\tt apec} model to the background model. 
The free parameters of the additional {\tt apec} are the temperature and emission measure. 
We fix the metal abundance to $Z_\sun$, because it is not constrained with a large statistical error if it is allowed to vary.
The additional {\tt apec} is subject to interstellar absorption ({\tt TBabs}) whose column density ($N_\mathrm{H,src}$) is independent of $N_\mathrm{H,bgd}$.
The background absorption column ($N_\mathrm{H,bgd}$) and normalization ($f_\mathrm{bgd}$) are also allowed to vary.

The fitting procedure is the same as in the background modeling but each sub-region is independently fitted. The best-fit parameters are summarized in \Tab{tab:srcfit}, and the model curves are shown in \Fig{fig:src_spectra}.
All spectra are reproduced by the model with $\chi^2_{\nu} \sim 1.1$.
If we add a model systematic error of 4\% to mimic instrumental calibration uncertainty, the fits are formally acceptable with a null hypothesis probability of $>$5\%.
We also try to fit non-equilibrium ionization plasma models ({\tt nei} and {\tt rnei} in Xspec), but they give no significant improvements.


\begin{deluxetable*}{cccccccc}
\tabletypesize{\small}
\tablecaption{X-ray plume model parameters. \label{tab:srcfit}}
\tablehead{
\nocolhead{region} 
& \colhead{$N_\mathrm{H,bgd}$} 
& \colhead{$N_\mathrm{H,src}$} 
& \colhead{$kT$} 
& \colhead{EM} 
& \colhead{$f_\mathrm{bgd}$} 
& \colhead{$\chi^2/\nu$} 
& \colhead{\makecell{Unabsorbed surface brightness\\in 1.5--3.0 keV}} 
\\
\nocolhead{reg}
& \colhead{($10^{22}$~cm$^{-2}$)}
& \colhead{($10^{22}$~cm$^{-2}$)}
& \colhead{(keV)}
& \colhead{(cm$^{-6}$~pc)}
& \nocolhead{f}
& \nocolhead{chi}
& \colhead{(erg~cm$^{-2}$~s$^{-1}$~arcmin$^{-1}$)}
}
\startdata
S1 & ${2.17}^{+0.09}_{-0.09}$ & ${1.92}^{+0.13}_{-0.08}$ & ${0.74}^{+0.02}_{-0.02}$ &  ${0.16}^{+0.01}_{-0.01}$ & ${1.42}^{+0.02}_{-0.02}$ & 1973.6/2021 & 0.71$\times 10^{-14}$\\
S2 & ${2.83}^{+0.11}_{-0.11}$ & ${2.93}^{+0.08}_{-0.06}$ & ${0.62}^{+0.01}_{-0.01}$ &  ${0.76}^{+0.03}_{-0.03}$ & ${1.52}^{+0.02}_{-0.02}$ & 2551.2/2449 & 2.47$\times 10^{-14}$\\
S3 & ${3.43}^{+0.15}_{-0.14}$ & ${3.41}^{+0.11}_{-0.08}$ & ${0.73}^{+0.01}_{-0.01}$ &  ${0.76}^{+0.03}_{-0.03}$ & ${1.92}^{+0.02}_{-0.02}$ & 2664.5/2410 & 3.27$\times 10^{-14}$\\
S4 & ${5.06}^{+0.46}_{-0.40}$ & ${3.65}^{+0.10}_{-0.09}$ & ${0.85}^{+0.03}_{-0.03}$ &  ${0.82}^{+0.09}_{-0.08}$ & ${2.63}^{+0.07}_{-0.07}$ & 824.2/765 & 4.37$\times 10^{-14}$\\
S5 & ${7.80}^{+0.52}_{-0.44}$ & ${4.26}^{+0.11}_{-0.11}$ & ${0.76}^{+0.03}_{-0.03}$ &  ${0.76}^{+0.08}_{-0.07}$ & ${5.22}^{+0.09}_{-0.08}$ & 1845.2/1747 & 3.51$\times 10^{-14}$\\
\enddata 
\tablecomments{Metalicity is fixed to the solar value.}
\end{deluxetable*}

\section{Disucssion}
\label{sec:dis}
\subsection{Distance and Physical Properties}
The measured absorption column densities for the X-ray plume ($N_\mathrm{H, src}$) are consistent with that of the background emission  ($N_\mathrm{H, bgd}$) in region S1, S2, and S3, indicating that the X-ray plume is not a foreground source but is in the GC region.
In the lower latitude S4 and S5 regions, $N_\mathrm{H, src}$ are still as large as a typical column density toward the GC region (several $10^{22}$ cm$^{-2}$), but are smaller than $N_\mathrm{H, bgd}$ by a factor of 1.4--1.9.
This can be explained by the inhomogeneous distribution of dense molecular clouds at $|b| \lesssim 0\fdg5$, the so-called Central Molecular Zone \citep[CMZ;][]{1996ARA&A..34..645M}.
Even though the line-of-sight distribution of the CMZ at this latitude is uncertain, our observations suggest that the molecular clouds could be located slightly to the far side of the GC and/or the X-ray plume could be located slightly to the near side of the GC.

Assuming a distance of 8~kpc to the X-ray plume, the projected physical size is estimated to be $\sim$100~pc $\times$ 50~pc. 
The line-of-sight depth is uncertain, but we assume 50~pc hereafter.
Accordingly, the hot gas density can be estimated from the observed EM to be 0.06--0.13~cm$^{-3}$. 
The thermal pressure of the gas is in the range (0.7--1.8)$\times10^{-10}$~erg~cm$^{-3}$.
The total mass is $\sim$600~$M_\sun$ and the total thermal energy reaches $7\times10^{50}$~erg.
Since the sound velocity of a $kT = 0.7$~keV gas is $\sim$$400$~km~s$^{-1}$, the sound crossing time of the X-ray plume is $2\times10^{5}$~yr, which gives a rough estimate of its age assuming expansion of the hot gas.



\subsection{Association with Other Wavelength}

\Fig{fig:image}(d) is a composite image of the radio continuum (red), polarized radio intensity (green), and $^{12}$CO clouds (blue).
The X-ray plume is to the east of the GC radio lobe \citep{1984Natur.310..568S}, which is considered to be a relic of a mass outflow from the GC $\approx$10~Myr ago.
No significant X-ray excesses over the diffuse background are found inside or to the west of the radio lobe. 
The polarized radio plume was found on the eastern edge of the radio lobe \citep{1985Natur.317..697S,1986AJ.....92..818T} and apparently connects the X-ray plume with the intense X-ray emission in the GC. 
The tower-like molecular cloud extending from $b\sim0\fdg1$ to $b\sim0\fdg5$ is a molecular counterpart to the double helix nebula discovered by \cite{2014ApJ...780...72E}. The cloud is located at the western edge of the polarized radio lobe.

\subsection{Possible Origin}
The base of the X-ray plume, corresponding to region S5 in \Fig{fig:image}(a), was reported by \cite{2015MNRAS.453..172P} and \cite{2018PASJ...70...82Y}, who named the structure the eastern edge and NE, respectively.
The physical properties they reported roughly agree with ours.
They also reported an X-ray excess on the western side, ($l$, $b$) $\sim$ ($359\fdg7$, $0\fdg2$), which is referred to as the western edge or NW, respectively.
They discussed the origin as multiple supernovae in the GC and/or past flares of Sgr A* in relation to the almost symmetric X-ray morphology of the east and west sides at $0\fdg1 < b < 0\fdg3$.
However, our extended survey reveals a highly asymmetric morphology at $0\fdg3 < b < 1\fdg0$, which raises an issue with this scenario; an outflow on the western side should be significantly suppressed if explosive events at the GC are the origin.
Given the existing outflow feature on the western side, i.e., the radio lobe (see \Fig{fig:image}(d)), there is no strong evidence for the hot gas outflow to be suppressed in this region.

Since the total thermal energy of the X-ray plume can be explained by a single supernova explosion, it might be a supernovae remnant (SNR) at high latitude.
Indeed, the temperature and abundance of the plume are similar to those of middle-aged Galactic SNRs \citep[e.g.,][]{2005ApJ...631..935K}.
However, its physical size is significantly larger than that of a typical SNR.
In addition, no corresponding radio emitting shell has been found \citep{2000AJ....119..207L,2008ApJS..177..255L}.
Therefore, an in-situ supernova is unlikely to be the origin of the X-ray plume, even though it cannot be completely ruled out.

As pointed out by \cite{2015MNRAS.453..172P}, the most characteristic feature of the X-ray plume is its apparent association with the polarized radio plume.
Therefore, the magnetic flux tube at the polarized radio plume likely plays a crucial role in forming the X-ray plume;
if the hot gas has a source at the footpoint of the radio plume, it would blow out along the magnetic flux tube.
Indeed, the magnetic field at the radio plume is estimated to be several tens of $\mu$G \citep{1986AJ.....92..818T},  giving a magnetic pressure comparable to the thermal pressure of the X-ray plume, $B^2/8\pi \approx 10^{-10}$~erg~cm$^{-3}$.
At the northern end of the radio plume, the polarized radio intensity (and hence the magnetic field) becomes one order of magnitude lower than the peak value, and the X-ray-emitting hot gas is no longer confined by the magnetic flux tube.
Besides, an average pressure of the cold interstellar medium at $b > 0.3\arcdeg$ is estimated to be $\lesssim$10$^{-11}$~erg~cm$^{-3}$ according to its spatial distribution \citep{2007A&amp;A...467..611F}, and is significantly lower than that of the X-ray plume.
This is because the observed X-rays extend to the north and east (even southeast) of the radio plume.
Even though the X-ray emission does not extend to the west of the radio plume, it can still be explained by external pressure owing to the warm ionized gas at the edge of the radio lobe \citep[$\sim$$5\times10^{-10}$~erg~cm$^{-3}$;][]{2009ApJ...695.1070L} and/or the tower cloud as shown in Figure 1(d).

If the above magnetized outflow scenario is the case, what is the source of the hot gas?
The current X-ray observations are inconclusive, but magnetic reconnection at the radio arc \citep{1994ApJ...424L..91S} or stellar winds from the Arches cluster \citep{2007PASJ...59S.229T} are viable candidates because of their proximity to the polarized radio lobe.
The arc bubble \citep{2013MNRAS.434.1339H,2015MNRAS.453..172P} is another possible source but is rather young (several 10$^4$ yr) to provide hot gas to the X-ray plume, whose sound crossing timescale is 0.2~Myr.

Finally, we comment relationship of the X-ray plume with the elongated hot gas at ($l\sim0$, $-1\fdg8<b<-1\fdg0$) found by \cite{2013ApJ...773...20N}.
This southern hot gas might also originate from the magnetized outflow, because the southern part of the polarized radio plume reaches $b\sim -0\fdg9$ \citep{1986AJ.....92..818T} and is connected with the hot gas.
The timescale of the southern hot gas ($\sim$0.1~Myr) is consistent with that of the northern X-ray plume.
However, its temperature is lower and its ionization state is different from that of the X-ray plume.
Determining the physical association between the northern and southern X-ray-emitting hot gases will require further investigations, especially at a low latitude region ($-1\fdg0 < b < -0\fdg2$).

\section{Conclusions}

From the X-ray survey of the northern GC with Suzaku, we discovered a plume-like structure around ($l$, $b$) $\sim$ (0\fdg2, 0\fdg6) with an angular size of $\sim$$1\arcdeg$.
The X-ray plume is located to the east of the radio lobe, and no significant X-ray excesses are found inside or to the west of the lobe. 
By carefully treating the diffuse background emission in the GC region, we reproduce the spectrum of the plume by using a single absorbed CIE plasma model with $kT \sim 0.7$~keV and the solar metal abundance.
The unabsorbed surface brightness in the 1.5--3.0~keV band averaged over the entire plume is $\sim$$3\times10^{-14}$~erg~cm$^{-2}$~s$^{-1}$~arcmin$^{-2}$.
The plume is likely located at the GC distance because of a large absorption column density of (2--4)$\times10^{22}$~cm$^{-2}$. 
We estimated an electron density of 0.1~cm$^{-3}$, thermal pressure of $1\times10^{-10}$~erg~cm$^{-3}$, total mass of 600~$M_\sun$, and total thermal energy of $7\times10^{50}$~erg.

The X-ray plume is spatially associated with the polarized radio plume, whose magnetic pressure is comparable to the thermal energy of the X-ray plume.
Therefore the most plausible origin of the X-ray plume is magnetized hot gas outflow from the GC.
External pressure due to the radio lobe and/or nearby molecular clouds likely affects its morphology.
The energy source could be magnetic reconnection at the radio arc or stellar winds from the Arches cluster.
The southern diffuse hot gas reported by \cite{2013ApJ...773...20N} could be physically associated with the X-ray plume, but further investigation is necessary to elucidate their true nature.

\acknowledgments
We thank M. Tsuboi and Y. Sofue for providing the polarized radio intensity image and valuable comments on the discussion.
This work is supported by JSPS KAKENHI Grant Number 16K17674 (S.N.).
S.N is also supported by the RIKEN SPDR Program.

\software{HEAsoft (v6.22; HEASARC 2014), XSPEC (v12.9.1t; Arnaud 1996)}

\bibliographystyle{aasjournal}
\bibliography{bibdesk}



\end{document}